\documentclass[preprint]{ptephy_v1}

\preprintnumber{XXXX-XXXX} 
\usepackage{hyperref}

\begin{document}

\title{Construction of the Temperley-Lieb algebra from bond algebra: From the transverse-field Ising to the XXZ}


\author{Yukihisa Imamura}
\affil{Yukawa Institute for Theoretical Physics, Kyoto University,
Kitashirakawa Oiwakecho, Sakyo-ku, Kyoto 606-8502 Japan \email{yukihisa.imamura@yukawa.kyoto-u.ac.jp}}





\begin{abstract}%
We show that the Temperley-Lieb algebra is constructed from the generators of the transverse-field Ising-type bond algebra.
It is also shown that, when we take the representation of the generators to the one-dimensional transverse-field Ising model, then the Temperley-Lieb algebra constructed on three consecutive bond generators becomes the XXZ model.
Furthermore, we obtain several integrable spin systems from other representations for the bond generators.
\end{abstract}

\subjectindex{A10, A43}

\maketitle

\section{Introduction}

Exactly solvable lattice spin models have been playing important roles in the understanding of physics in strongly correlated systems. Since the age of Onsager's remarkable work \cite{Onsager-1944}, quantum solvable lattice models have been used as toy models for understanding physical concepts and phenomena.
Quantum solvable lattice spin models are classified into three types.
The first one has a Hamiltonian of which terms commute with each other, such as the Kitaev toric code \cite{Kitaev-2003} and the X-cube code \cite{Casteinovo-Chamon-Sherrington-2010,Vijay-Haah-Fu-2015}. These are used in the field of quantum computation as stabilizer codes and quantum error corrections \cite{Nielsen-Chuang-2000-book}.
The second one has a Hamiltonian of which terms are transformed into Majorana-bilinears. For instance, the one-dimensional (1d) transverse-field Ising (1dTFI) model \cite{Pfeuty-1970} and the 1d XY model \cite{Lieb-Schultz-Mattis-1961} belong to this class, because the spin variables are converted to the product of Majorana fermions by using the Jordan-Wigner transformation \cite{Jordan-Wigner-1928}. The Kitaev's honeycomb lattice model \cite{Kitaev-2006} is also included in this class.
The third one is the class with a more complicated but beautiful structure, the quantum integrability. This type includes the 1d Heisenberg model \cite{Heisenberg-1928} and the XXZ model \cite{Yang-Yang-1966-1,Yang-Yang-1966-2,Yang-Yang-1966-3,Yang-Yang-1966-4}. These models can be solved by using the Bethe ansatz \cite{Bethe-1931,Faddeev-1996}, and have the remarkable algebraic relation, the Yang-Baxter equation \cite{Faddeev-1996}, the Temperley-Lieb (TL) algebra \cite{Temperley-Lieb-1971}.
The TL algebra has been historically introduced to construct  transfer matrices of two-dimensional statistical physics models \cite{Temperley-Lieb-1971}, and now is related to quantum integrable models, the quantum groups, and the braid groups.
Well-known examples are the XXZ model and the Potts model for one-dimensional quantum systems and the six-vertex model for two-dimensional classical systems \cite{Baxter-book}.

\begin{figure}
\centering
    \includegraphics[width=12cm]{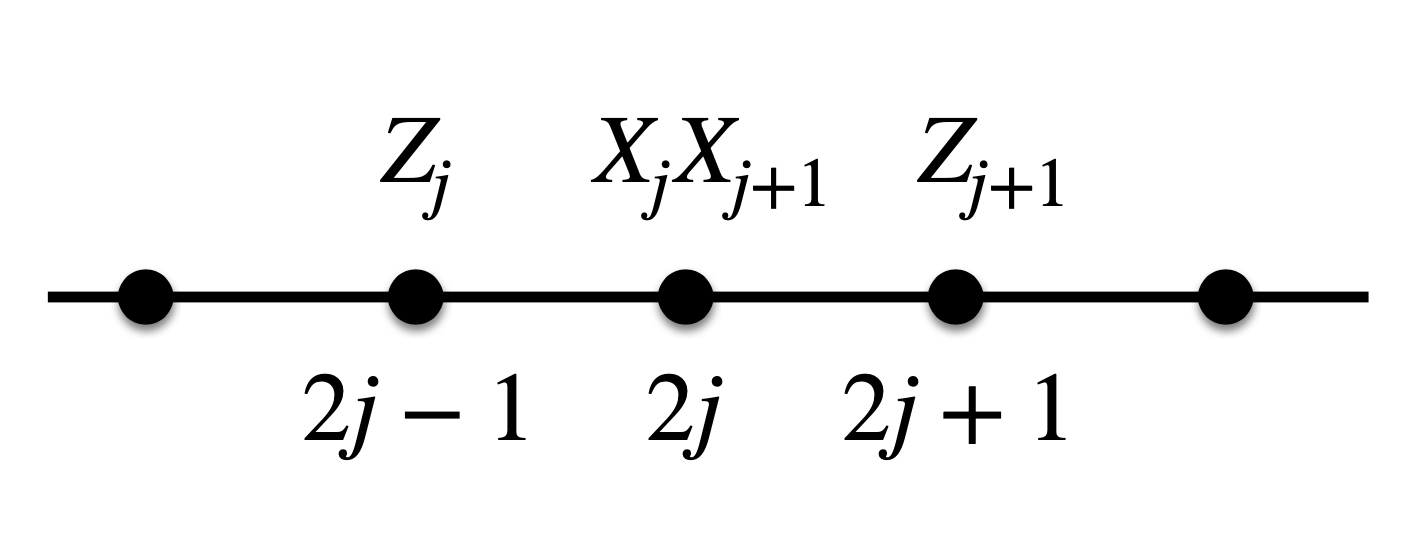}
    \caption{The graph of the 1dTFI bond algebra. The Hamiltonian is represented by the Ising terms $X_jX_{j+1}$ and the transverse-field terms $Z_j$, so $h_{2j-1}$ is $Z_j$ and $h_{2j}$ is $X_jX_{j+1}$.
    We associate each $h_j$ with a vertex and draw the graph.}
    \label{fig:1dTFI}
\end{figure}

Recently, there exists progress in the second class.
Assuming that a Hamiltonian has the form $H=\sum_j \lambda_j h_j$, we define an algebra generated from $\{h_j\}$ \cite{Nussinov-Ortiz-2009}.
For example, the 1dTFI case is shown in Fig.\ref{fig:1dTFI}.
We call it the bond algebra generated from $\{h_j\}$.
Using this algebra, it has been found that the usual Jordan-Wigner transformation is algebraically generalized by a few algebraic rules of the generators \cite{Minami-2016}.
The rules can be represented as a graph, which includes vertices and lines \cite{Ogura-Imamura-Kameyama-Minami-Sato-2020}.
By considering shapes of graphs, we have found several lattice models, whose Hamiltonians take the form of Majorana-bilinears.
This approach enables us to treat models such as the 1dTFI model and the Kitaev's honeycomb model \cite{Kitaev-2006,Kitaev-Laumann-2009,Minami-2019}, in a systematic and unified way.
We call this method the bond algebraic approach.


\begin{figure}
\centering
    \includegraphics[width=12cm]{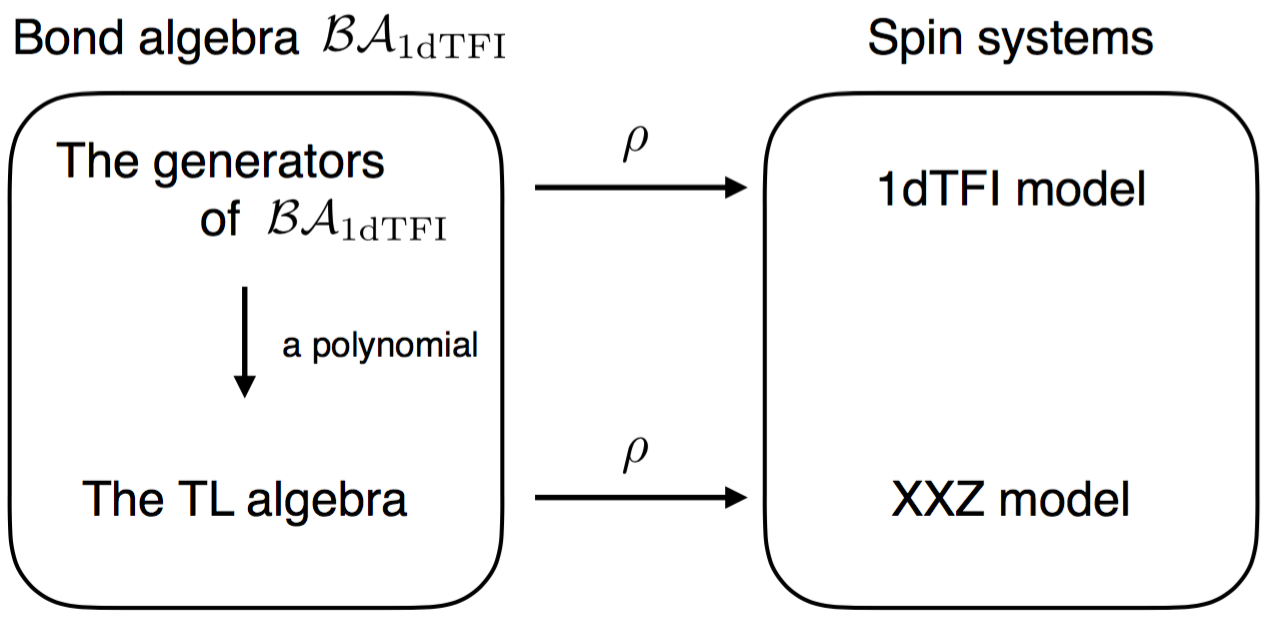}
    \caption{Conceptual diagram relating the 1dTFI bond algebra and spin systems. The Temperley-Lieb (TL) algebra is obtained as a polynomial of the bond generators. The representation $\rho$ maps the generators into the 1dTFI model and the TL algebra into the XXZ model.}
    \label{fig:TFI-TL-concept}
\end{figure}

In this paper, we attempt to connect the second and the third classes of solvable spin systems.
We prove that the TL algebra can be constructed as a set of polynomials of the bond generators, and namely is included in the bond algebra.
In previous studies \cite{Nussinov-Ortiz-2009,Minami-2016,Minami-2019,Ogura-Imamura-Kameyama-Minami-Sato-2020}, we have mainly considered a set of bond generators; it defines a Hamiltonian of the second class.
However, a bond algebra consists of polynomials of the bond generators as well as the generators themselves, and therefore it has a more fruitful structure.
We focus on this point and consider whether the TL algebra is included in a bond algebra.
We prove that the TL algebra is constructed from the 1dTFI generators.
Specifically, we try to construct the generators of the TL algebra, separately for cases of one, two, and three consecutive vertices of the 1dTFI bond algebra.

This paper is organized as follows.
In Sec. \ref{sec:preliminaries}, we establish the notations and review the bond algebra method for the 1dTFI model and the TL algebra.
In Sec. \ref{sec:TL-on-1and2}, we show that the TL algebra can be constructed in the 1dTFI bond algebra for cases of one and two consecutive vertices.
We derive the conditions whether the TL algebra can be constructed.
In Sec. \ref{sec:TL-on-3}, we show that the TL algebra can be constructed on three consecutive vertices, which is our main result.
Moreover, it is found that when the 1dTFI bond algebra is represented by the 1d TFI model, the TL generators coincide with the Hamiltonian of the XXZ model.
Furthermore, we give several other representations of the 1dTFI bond algebra and the spin systems which correspond to the representations.
We conclude with a discussion of future directions in Sec. \ref{sec:conclusion}.

\section{Preliminaries}
\label{sec:preliminaries}
We start with a review of preliminaries.
In Sec. \ref{subsec:TFIM-BA}, we explain the bond algebra of the 1dTFI model \cite{Minami-2016,Ogura-Imamura-Kameyama-Minami-Sato-2020}.
In Sec. \ref{subsec:TL}, the definition and the notations of the TL algebra are established.

\subsection{The 1d transverse-field Ising model and its bond algebra}
\label{subsec:TFIM-BA}

\begin{figure}
    \centering
    \includegraphics[width=12cm]{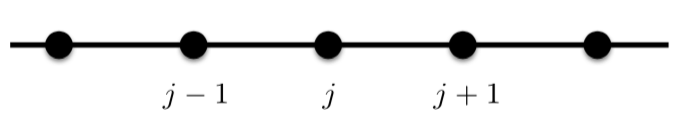}
    \caption{The commutativity graph of the 1dTFI bond algebra. Each vertex represents a generator, and each line between them represents the anti-commutation relation.}
    \label{fig:1dTFIBA}
\end{figure}

Let us start to consider the following Hamiltonian
\begin{equation}
    H=\sum_j \lambda_j h_j ~~(j\in\mathbb{Z}),
\end{equation}
where $\lambda_j$ are real parameters and the operators $\{h_j\}$ satisfy the following algebraic relations:
\begin{equation}
\label{eq:1dTFI-BA}
    h_j^2=1,~~
    h_j h_{j+1} = - h_{j+1} h_j, ~~
    h_j h_k= h_k h_k ~~ (|j-k|>1).
\end{equation}
The bond algebra $\mathcal{BA}_{\mathrm{1dTFI}}$ is defined as the algebra generated from these $\{h_j\}$ \cite{Nussinov-Ortiz-2009,Minami-2016}, which we call the 1dTFI bond algebra.
We associate each $\mathcal{BA}$ with a graph $\mathcal{G}(\mathcal{BA})$ by the following rules \cite{Ogura-Imamura-Kameyama-Minami-Sato-2020};
\begin{itemize}
    \item $h_j$ is vertex and labelled $j$-th.
    \item When $h_j$ and $h_k$ anti-commute (commute) with each other, we draw (do not draw) a line between the vertices with $j$-th and $k$-th.
\end{itemize}
We call $\mathcal{G}(\mathcal{BA})$ the commutativity graph of $\mathcal{BA}$.
The commutativity graph of the 1dTFI bond algebra $\mathcal{G}(\mathcal{BA}_{\mathrm{1dTFI}})$ is shown in Fig.\ref{fig:1dTFIBA}.

The Hamiltonian of the ordinary 1dTFI chain can be obtained by taking the following representation,
\begin{equation}
\label{eq:1dTFI-Rep1}
    \rho_{\mathrm{TFI}}\left(h_{2j-1}\right)=Z_j,~~\rho_{\mathrm{TFI}}\left(h_{2j}\right)=X_jX_{j+1}.
\end{equation}
Assume that the parameters are uniform, $\lambda_{2j-1}=\lambda_{\mathrm{odd}}$, $\lambda_{2j}=\lambda_{\mathrm{even}}$, then the Hamiltonian is represented as follows:
\begin{equation}
\label{eq:1dTHI-Hamiltonian}
    \rho_{\mathrm{TFI}}\left(H\right)=H_{\mathrm{1dTFI}}=\sum_j \left( \lambda_{\mathrm{odd}} Z_j + \lambda_{\mathrm{even}} X_jX_{j+1}\right),
\end{equation}
which is the ordinary 1dTFI chain.
Note that there is another representation for $\mathcal{BA}_{\mathrm{1dTFI}}$;
\begin{equation}
\label{eq:1dTFI-Rep2}
    \Tilde{\rho}_{\mathrm{TFI}}\left(h_{2j-1}\right)=X_{j-1}X_{j},~~\Tilde{\rho}_{\mathrm{TFI}}\left(h_{2j}\right)=Z_j.
\end{equation}
This representation leads to the (self-)dual Hamiltonian of $H_{\mathrm{1dTFI}}$,
\begin{equation}
    \Tilde{\rho}_{\mathrm{TFI}}\left(H\right)=\Tilde{H}_{\mathrm{1dTFI}}=\sum_j (\lambda_{\mathrm{odd}} X_{j-1}X_{j} + \lambda_{\mathrm{even}} Z_j).
\end{equation}

\subsection{The Temperley-Lieb algebra}
\label{subsec:TL}

The Temperley-Lieb (TL) algebra is generated by a sequence of generators, $e_j$ satisfying the following rules
\begin{align}
    &e_j^2=\Delta e_j,\label{eq:TL-projection}\\
    &e_je_{j\pm1}e_j=e_j,\label{eq:TL-gluing}\\
    &e_je_k=e_ke_j~~(|j-k|>1)\label{eq:TL-commutativity},
\end{align}
where $\Delta$ is a real parameter.

Let us consider the meanings of these rules.
The first one (\ref{eq:TL-projection}) shows the almost projectivity of the operators.
If we write $p_j=e_j/\Delta$, then $p_j$ satisfies the usual property of a projection operator; $p_j^2=p_j$.
The second one (\ref{eq:TL-gluing}) is, in this paper, often called the gluing conditions.
If all the generators commuted with each other, then the model 
\begin{equation}
    H=\sum_j e_j
\end{equation}
would be naturally solvable by optimizing the individual terms.
Despite the non-commutativity (\ref{eq:TL-gluing}) among the generators, the Hamiltonian is known solvable, or quantum integrable \cite{Martin-book,Jones-2012}.
This is a remarkable feature of the TL algebra.

The XXZ model is known to satisfy the TL algebra \cite{Martin-book}.
The TL generators are represented by
\begin{equation}
    e_j=
    -\frac{1}{2}\left(X_j X_{j+1}+Y_jY_{j+1}\right)-
    \frac{q+q^{-1}}{4}\left(Z_jZ_{j+1}-1\right)+
    \frac{q-q^{-1}}{4}\left(Z_j-Z_{j+1}\right),
\end{equation}
where $q$ is a complex parameter satisfying $q+q^{-1}=\Delta$.

\section{The TL algebra on one and two consecutive vertices}
\label{sec:TL-on-1and2}

Before proceeding to our main result given in Sec.\ref{sec:TL-on-3}, let us solve a few simpler problems to understand the TL algebra
included in the 1dTFI-type bond algebra $\mathcal{BA_{\mathrm{1dTFI}}}$.
We attempt to represent the TL generators $e_j$ as a polynomial of the generators $\{h_j\}$ on several vertices around $j$:
\begin{equation}
    e_j=e_j(\dots,h_{j-1},h_j,h_{j+1},\dots).
\end{equation}

\subsection{The TL generator on a single vertex}
\label{subsec:TL-on-1}
Let us put a projection operator on $j$-th vertex as 
\begin{equation}
    p_j=\frac{1}{2}(x+yh_j),
\end{equation}
where $x$ and $y$ are real parameters.
Using (\ref{eq:1dTFI-BA}), we calculate $p_j^2$ as
    $p_j^2 = ((x^2+y^2) + 2xyh_j)/4$.
For the projectivity condition $p_j^2=p_j$, $x$ and $y$ need to satisfy
\begin{equation}
    x=1,~~y^2=1,
\end{equation}
and therefore the projection operator on the $j$-th vertex has the form
\begin{equation}
    p_j=\frac{1}{2}(1+h_j), ~~1-p_j=\frac{1}{2}(1-h_j).
\end{equation}
Here we choose $y=1$ for $p_j$ (automatically $y=-1$ corresponding to the complementary projection $1-p_j$).

Now let us consider the construction of the TL generators satisfying the algebraic relations (\ref{eq:TL-projection}), (\ref{eq:TL-commutativity}), and
(\ref{eq:TL-gluing}).
For the TL generators to satisfy the commutation relations (\ref{eq:TL-commutativity}), we define $e_j=\Delta p_j$. The commutation relations $e_je_{j+2}=e_{j+2}e_j$ is automatically satisfied from the commutativity of the bond generators $h_jh_{j+2}=h_{j+2}h_j$.
Next let us determine $\Delta$ by the gluing condition (\ref{eq:TL-gluing}).
From the anti-commutation relation of the generators of $\mathcal{BA}_{\mathrm{1dTFI}}$
\begin{equation}
\begin{split}
p_j h_{j\pm1} p_j
&= h_{j\pm 1}\frac{1}{2}(1-h_j)p_j\\
&= h_{j\pm 1}(1-p_j)p_j\\
&=0,
\end{split}
\end{equation}
and then the gluing condition (\ref{eq:TL-gluing}) is calculated as
\begin{equation}
    e_j e_{j\pm 1}e_j
    =\frac{\Delta}{2}e_j^2
    =\frac{\Delta^2}{2}e_j,
\end{equation}
which means $\Delta^2=2$.
Thus we construct the TL generator on the $j$-th vertex,
\begin{equation}
    e_j=\frac{\Delta}{2}(1+h_j),~~\Delta^2=2.
\end{equation}
When we represent the generators $h_j$ as (\ref{eq:1dTFI-Rep1}), $e_j$ becomes
\begin{equation}
    e_{2j-1}=\frac{\Delta}{2}(1+Z_j),~~ e_{2j}=\frac{\Delta}{2}(1+X_jX_{j+1}).
\end{equation}
This is equivalent to the 2-state Potts model.

\subsection{The TL generators on two consecutive vertices}
\label{subsec:TL-on-2}

A generic projection operator containing $h_j$ and $h_{j+1}$ on the two consecutive vertices $j$ and $(j+1)$ takes the following form: 
\begin{equation}
    p_{j+\frac{1}{2}}=\frac{1}{2}(1+xh_j + yh_{j+1} + z\mathrm{i} h_j h_{j+1}),
\end{equation}
where $x,y,z$ are real parameters.
Using (\ref{eq:1dTFI-BA}), $p_{j+\frac{1}{2}}^2$ is calculated as
\begin{equation}
    p_{j+\frac{1}{2}}^2=\frac{1}{4}\left((1+x^2+y^2+z^2) + 2xh_j + 2yh_{j+1} + 2z\mathrm{i} h_j h_{j+1}\right).
\end{equation}
In order for the projectivity condition $p_{j+\frac{1}{2}}^2=p_{j+\frac{1}{2}}$ to hold, the parameters $x,y,z$ need to satisfy
 $x^2+y^2+z^2=1$,
and the projection operator on $j$-th vertex has the form
\begin{equation}
\label{eq:proj-on-2}
\begin{split}
&p_{j+\frac{1}{2}}=\frac{1}{2}(1+xh_j + yh_{j+1} + z\mathrm{i} h_j h_{j+1}),\\
&\text{with} ~~x^2+y^2+z^2=1.
\end{split}
\end{equation}
We note that the inversion on the sphere $(x,y,z) \to (-x,-y,-z)$ leads to the complementary projection $1-p_{j+\frac{1}{2}}$.
\begin{figure}
    \centering
    \includegraphics[width=10cm]{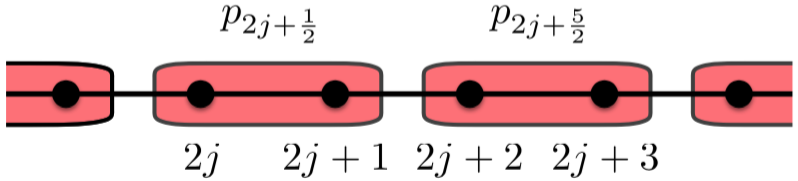}
    \caption{Projection operators defined on two consecutive vertices.}
    \label{fig:TL-2}
\end{figure}

Next, we want to consider the TL algebra, but unfortunately, they cannot be constructed in this case.
The reason is as follows.
In order for the TL generators to satisfy the commutation relations (\ref{eq:TL-commutativity}), i.e. $e_je_{j+2}=e_{j+2}e_j$, we need to put $e_j=\Delta p_{2j+\frac{1}{2}}$ instead of $e_j=\Delta p_{j+\frac{1}{2}}$ such as in Fig.\ref{fig:TL-2}.
Since
\begin{equation}
    \begin{split}
    p_{j+\frac{1}{2}}h_{j+2}p_{j+\frac{1}{2}}
    &=h_{j+2}\frac{1}{2}(1+x_jh_j-y_{j+1}h_{j+1}-z_{j+\frac{1}{2}}\mathrm{i}h_jh_{j+1})p_{j+\frac{1}{2}}\\
    &=h_{j+2}(1-p_{j+\frac{1}{2}}+x_jh_j)p_{j+\frac{1}{2}}\\
    &=x_jh_jh_{j+2}p_{j+\frac{1}{2}},
    \end{split}
\end{equation}
\begin{equation}
    p_{j+\frac{1}{2}}h_{j+3}p_{j+\frac{1}{2}}=h_{j+3}p_{j+\frac{1}{2}},
\end{equation}
\begin{align}
p_{j+\frac{1}{2}}\mathrm{i}h_{j+2}h_{j+3}p_{j+\frac{1}{2}}
=x_jh_j(\mathrm{i}h_{j+2}h_{j+3})p_{j+\frac{1}{2}},
\end{align}
the left hand side of the gluing condition (\ref{eq:TL-gluing}) can be calculated as
\begin{align}
e_j e_{j+1} e_j =\frac{\Delta^2}{2}(1+x_jx_{j+2}h_jh_{j+2}+y_{j+3}h_{j+3}+x_jz_{j+\frac{5}{2}}h_j(\mathrm{i}h_{j+2}h_{j+3}))e_j.
\end{align}
Any parameters can not eliminate the terms other than the first, and therefore it fails.

We note that, although we fail to construct the TL generators by $h_j$ on two consecutive vertices, the projection operators $p_{j+\frac{1}{2}}$ in (\ref{eq:proj-on-2}) can be used in a renormalization group for the bond algebra, which relates to Monthus's work \cite{Monthus-2015}.
This fact is not the main topic of this paper, so it is elaborated in \ref{appendix:RG}.


\section{The TL algebra on three consecutive vertices}
\label{sec:TL-on-3}

This section is the main part of this paper.
We construct the TL generators using the $\mathcal{BA}_{\mathrm{1dTFI}}$ generators $h_j$ on three consecutive vertices and find that they can be represented as the XXZ model.
In the course of the computations, we introduce ``pseudo'' Pauli operators.

\begin{figure}
    \centering
    \includegraphics[width=12cm]{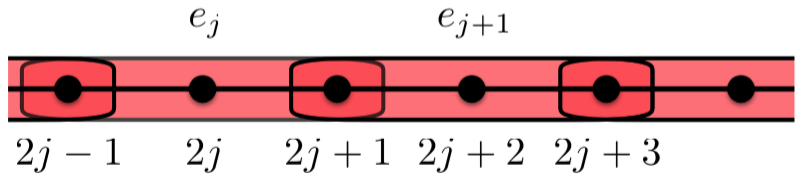}
    \caption{The TL algebra on three consecutive vertices.}
    \label{fig:TL-3}
\end{figure}

First of all let us try to find a projection operators consisting of the generators $\{h_j\}$ on the three consecutive vertices $j-1$, $j$ and $j+1$ (see Fig.\ref{fig:TL-3}).
A generic form is given as
\begin{align}
    p_j
    &=\frac{1}{4}\left[
    (w_{1,j}-w_{2,j}h_{j-1}h_{j+1})
    +(x_{1,j}h_j+x_{2,j}h_{j-1}h_{j}h_{j+1})\right.\nonumber\\
    &\hspace{25mm}\left.
    +(y_{1,j}\mathrm{i}h_{j-1}h_{j}+y_{2,j}\mathrm{i}h_{j}h_{j+1})
    +(z_{1,j}h_{j-1}-z_{2,j}h_{j+1})
    \right]\\
    &=\frac{1}{2}\left(
    \mathcal{I}_j(w_{1,j},w_{2,j})
    +\mathcal{X}_j(x_{1,j},x_{2,j})
    +\mathcal{Y}_j(y_{1,j},y_{2,j})
    +\mathcal{Z}_j(z_{1,j},z_{2,j})
    \right).
\end{align}
Here $(w_{n,j},x_{n,j},y_{n,j},z_{n,j})_{n=1,2}$ are real parameters.
Furthermore, we have introduced the ``pseudo" Pauli operators, $\mathcal{I}_j, \mathcal{X}_j, \mathcal{Y}_j, \mathcal{Z}_j$.
The reason why we call them ``pseudo" Pauli is that they satisfy the following relations:
\begin{align}
    &\left[\mathcal{I}_j, \mathcal{X}_j\right]
    =\left[\mathcal{I}_j, \mathcal{Y}_j\right]
    =\left[\mathcal{I}_j, \mathcal{X}_j\right]=0,\\
    &\{\mathcal{X}_j, \mathcal{Y}_j\}
    =\{\mathcal{Y}_j, \mathcal{Z}_j\}
    =\{\mathcal{Z}_j, \mathcal{X}_j\}=0,\\
    &{\mathcal{I}_j(w_{1,j},w_{2,j})}^2=\mathcal{I}_j\left(\frac{w_{1,j}^2+w_{2,j}^2}{2},w_{1,j}w_{2,j}\right)\\
    &{\mathcal{X}_j(x_{1,j},x_{2,j})}^2=\mathcal{I}_j\left(\frac{x_{1,j}^2+x_{2,j}^2}{2},x_{1,j}x_{2,j}\right)\\
    &{\mathcal{Y}_j(y_{1,j},y_{2,j})}^2=\mathcal{I}_j\left(\frac{y_{1,j}^2+y_{2,j}^2}{2},y_{1,j}y_{2,j}\right)\\
    &{\mathcal{Z}_j(z_{1,j},z_{2,j})}^2=\mathcal{I}_j\left(\frac{z_{1,j}^2+z_{2,j}^2}{2},z_{1,j}z_{2,j}\right)\\
    &\mathcal{I}_j(w_{1,j},w_{2,j})\mathcal{X}_j(x_{1,j},x_{2,j})=\mathcal{X}_j\left(\frac{w_{1,j}x_{1,j}+w_{2,j}x_{2,j}}{2},\frac{w_{1,j}x_{2,j}+w_{2,j}x_{1,j}}{2}\right)\\
    &\mathcal{I}_j(w_{1,j},w_{2,j})\mathcal{Y}_j(y_{1,j},y_{2,j})=\mathcal{Y}_j\left(\frac{w_{1,j}y_{1,j}+w_{2,j}y_{2,j}}{2},\frac{w_{1,j}y_{2,j}+w_{2,j}y_{1,j}}{2}\right)\\
    &\mathcal{I}_j(w_{1,j},w_{2,j})\mathcal{Z}_j(z_{1,j},z_{2,j})=\mathcal{X}_j\left(\frac{w_{1,j}z_{1,j}+w_{2,j}z_{2,j}}{2},\frac{w_{1,j}z_{2,j}+w_{2,j}z_{1,j}}{2}\right).
\end{align}

Using these fundamental equations $p_j^2$ is computed as follows:
\begin{equation}
    \begin{split}
    p_j^2
    &=\frac{1}{4}\left[
    {\mathcal{I}_j(w_{1,j},w_{2,j})}^2
    +{\mathcal{X}_j(x_{1,j},x_{2,j})}^2
    +{\mathcal{Y}_j(y_{1,j},y_{2,j})}^2
    +{\mathcal{Z}_j(z_{1,j},z_{2,j})}^2\right.\\
    &\left.\hspace{5mm}+2\mathcal{I}_j(w_{1,j},w_{2,j})\mathcal{X}_j(x_{1,j},x_{2,j})
    +2\mathcal{I}_j(w_{1,j},w_{2,j})\mathcal{Y}_j(y_{1,j},y_{2,j})
    +2\mathcal{I}_j(w_{1,j},w_{2,j})\mathcal{Z}_j(z_{1,j},z_{2,j})
    \right]\\
    &=\frac{1}{2}\left[
    \mathcal{I}_j\left(\frac{1}{4}\sum_{n=1,2}(w_{n,j}^2+x_{n,j}^2+y_{n,j}^2+z_{n,j}^2),\frac{1}{2}(w_{1,j}w_{2,j}+x_{1,j}x_{2,j}+y_{1,j}y_{2,j}+z_{1,j}z_{2,j})\right)\right.\\
    &\hspace{10mm}
    +\mathcal{X}_j\left(\frac{w_{1,j}x_{1,j}+w_{2,j}x_{2,j}}{2},\frac{w_{1,j}x_{2,j}+w_{2,j}x_{1,j}}{2}\right)
    +\mathcal{Y}_j\left(\frac{w_{1,j}y_{1,j}+w_{2,j}y_{2,j}}{2},\frac{w_{1,j}y_{2,j}+w_{2,j}y_{1,j}}{2}\right)\\
    &\hspace{50mm}\left.
    +\mathcal{Z}_j\left(\frac{w_{1,j}z_{1,j}+w_{2,j}z_{2,j}}{2},\frac{w_{1,j}z_{2,j}+w_{2,j}z_{1,j}}{2}\right)
    \right].
    \end{split}
\end{equation}
For the projectivity $p_j^2=p_j$, the following relations must be satisfied:
\begin{align}
    &\frac{1}{4}\sum_{n=1,2}(w_{n,j}^2+x_{n,j}^2+y_{n,j}^2+z_{n,j}^2)=w_{1,j},\nonumber\\
    &\hspace{10mm}\frac{1}{2}(w_{1,j}w_{2,j}+x_{1,j}x_{2,j}+y_{1,j}y_{2,j}+z_{1,j}z_{2,j})=w_{2,j},\label{eq:wxyz}\\
    &\frac{1}{2}(w_{1,j}x_{1,j}+w_{2,j}x_{2,j})=x_{1,j},~~\frac{1}{2}(w_{1,j}x_{2,j}+w_{2,j}x_{1,j})=x_{2,j},\label{eq:wx}\\
    &\frac{1}{2}(w_{1,j}y_{1,j}+w_{2,j}y_{2,j})=y_{1,j},~~\frac{1}{2}(w_{1,j}y_{2,j}+w_{2,j}y_{1,j})=y_{2,j},\label{eq:wy}\\
    &\frac{1}{2}(w_{1,j}z_{1,j}+w_{2,j}z_{2,j})=z_{1,j},~~\frac{1}{2}(w_{1,j}z_{2,j}+w_{2,j}z_{1,j})=z_{2,j}\label{eq:wz}.
\end{align}
From (\ref{eq:wx}), we find that
\begin{equation}
    (x_{1,j}-x_{2,j})(w_{1,j}-w_{2,j}-2)=0,~~(x_{1,j}+x_{2,j})(w_{1,j}+w_{2,j}-2)=0,
\end{equation}
and therefore the allowed combinations are
\begin{equation}
    (w_{1,j},w_{2,j},x_{1,j},x_{2,j})=(2-w_j,w_j,x_j,x_j)~~ \text{or}~~(2-w_j,-w_j,x_j,-x_j)\label{eq:wx2},
\end{equation}
where $w$ and $x$ are real parameters.
Repeating the same steps for (\ref{eq:wy}), and (\ref{eq:wz}), we finally obtain the following two possibilities:
\begin{align}
    &(w_{1,j}, w_{2,j}, x_{1,j},x_{2,j}, y_{1,j},y_{2,j},z_{1,j},z_{2,j})\nonumber\\
    &\hspace{10mm}=(2-w_j,w_j,x_j,x_j,y_j,y_j,z_j,z_j)\label{eq:wxyz-1}\\
    &\hspace{15mm} \text{or}~~(2-w_j,-w_j,x_j,-x_j,y_j,-y_j,z_j,-z_j)\label{eq:wxyz-2}
\end{align}
where $y_j$ and $z_j$ are also real parameters.
Using (\ref{eq:wxyz-1}), or (\ref{eq:wxyz-2}), from (\ref{eq:wxyz}), we obtain the following condition among $w,x,y,z$
\begin{equation}
    w_j^2=1,~~ x_j^2+y_j^2+z_j^2=1.
\end{equation}
Thus, the allowed combinations are divided into four cases:
\begin{align}
    (w_{1,j},w_{2,j},x_{1,j},x_{2,j},y_{1,j},y_{2,j},z_{1,j},z_{2,j})
    &=(1,1,x_j,x_j,y_j,y_j,z_j,z_j)\\
    &\text{or}~~ (3,-1,x_j,x_j,y_j,y_j,z_j,z_j)\\
    &\text{or}~~ (1,-1,x_j,-x_j,y_j,-y_j,z_j,-z_j)\\
    &\text{or}~~ (3,1,x_j,-x_j,y_j,-y_j,z_j,-z_j).
\end{align}
Below we consider only the first $(w_{1,j}=w_{2,j}=1)$ and the third $(w_{1,j}=-w_{2,j}=1)$ ones:
\begin{equation}
\label{eq:projector-on-3}
\begin{split}
    p_j
    &=\frac{1}{4}\left[
    (1-h_{j-1}h_{j+1})
    +x_j(h_j+h_{j-1}h_{j}h_{j+1})
    +y_j(\mathrm{i}h_{j-1}h_{j}+\mathrm{i}h_{j}h_{j+1})
    +z_j(h_{j-1}-h_{j+1})
    \right]\\
    &=\frac{1}{2}(\mathcal{I}_j+x_j\mathcal{X}_j +y_j\mathcal{Y}_j+z_j\mathcal{Z}_j),
\end{split}
\end{equation}
\begin{equation}
\label{eq:projector-on-3-conj}
\begin{split}
    \bar{p}_j
    &=\frac{1}{4}\left[
    (1+h_{j-1}h_{j+1})
    +x_j(h_j-h_{j-1}h_{j}h_{j+1})
    +y_j(\mathrm{i}h_{j-1}h_{j}-\mathrm{i}h_{j}h_{j+1})
    +z_j(h_{j-1}+h_{j+1})
    \right]\\
    &=\frac{1}{2}(\bar{\mathcal{I}}_j+x_j\bar{\mathcal{X}}_j +y_j\bar{\mathcal{Y}}_j+z_j\bar{\mathcal{Z}}_j),
\end{split}
\end{equation}
because the others are obtained from the two by replacing $p_j \rightarrow 1-p_j$, $\bar{p}_j \rightarrow 1-\bar{p}_j$ and inverting the point $(x,y,z)$ into the opposite one $(-x,-y,-z)$.
Here we introduce the following simplified notations:
\begin{align}
    &\mathcal{I}_j=\mathcal{I}_j(1,1),~~\mathcal{X}_j=\mathcal{X}_j(1,1),~~
    \mathcal{Y}_j=\mathcal{Y}_j(1,1),~~\mathcal{Z}_j=\mathcal{Z}_j(1,1),\label{eq:psuedoPauli}\\
    &\bar{\mathcal{I}}_j=\mathcal{I}_j(1,-1),~~\bar{\mathcal{X}}_j=\mathcal{X}_j(1,-1),~~
    \bar{\mathcal{Y}}_j=\mathcal{Y}_j(1,-1),~~\bar{\mathcal{Z}}_j=\mathcal{Z}_j(1,-1).\label{eq:pseudoPauliconjugate}
\end{align}
The remarkable feature is that any products of the ``pseudo" Pauli operators (\ref{eq:psuedoPauli}) and the conjugates (\ref{eq:pseudoPauliconjugate}) equal zero; for instance,
\begin{equation}
    \begin{split}
    \mathcal{I}_j\bar{\mathcal{X}}_j
    &=\frac{1}{4}(1-h_{j-1}h_{j+1})(h_j-h_{j-1}h_{j}h_{j+1})\\
    &=\frac{1}{4}(h_j-h_{j-1}h_{j}h_{j+1}+h_{j-1}h_{j}h_{j+1}-h_j)\\
    &=0.
    \end{split}
\end{equation}

Next, let us consider the TL gluing condition (\ref{eq:TL-gluing}).
The TL generators in the blocking shown in Fig.\ref{fig:TL-3} are defined by $e_j=\Delta p_{2j}$ with $p_{2j}$ given by (\ref{eq:projector-on-3}).
The reason why $e_j$ is set to $\Delta p_{2j}$ is that they satisfy the non-commutative relations $[e_j, e_{j\pm1}]\neq0$, and the commutative relations $[e_j, e_k]=0~(|j-k|>1)$.
Using the ``pseudo" Pauli operators, we can calculate the individual terms in $e_j e_{j+1} e_j$ easily; for example,
\begin{equation}
    \begin{split}
    p_jh_{j+1}h_{j+3}p_j
    &=h_{j+1}h_{j+3}\frac{1}{2}\left(\mathcal{I}_j-x_j\mathcal{X}_j-y_j\mathcal{Y}_j+z_j\mathcal{Z}_j\right)p_j\\
    &=h_{j+1}h_{j+3}\left(\mathcal{I}_j-p_j+z_j\mathcal{Z}_j\right)p_j\\
    &=h_{j+1}h_{j+3}\left(-\bar{\mathcal{I}}_j+(1-p_j)+z_j\mathcal{Z}_j\right)p_j\\
    &=z_jh_{j+1}h_{j+3}\mathcal{Z}_j p_j\\
    &=-z_jh_{j+3}\mathcal{I}_j p_j\\
    &=-z_jh_{j+3} p_j.
    \end{split}
\end{equation}
Similarly we find that
\begin{align}
    p_j1p_j=p_j, ~~p_jh_{j+1}p_j=-z_jp_j, ~~p_jh_{j+3}p_j=h_{j+3}p_j.
\end{align}
Fortunately, the other terms vanish thanks to the fact that the products of the pseudo Pauli operators $(\mathcal{I}_j,\mathcal{X}_j,\mathcal{Y}_j,\mathcal{Z}_j)$ and their conjugations $(\bar{\mathcal{I}}_j,\bar{\mathcal{X}}_j,\bar{\mathcal{Y}}_j,\bar{\mathcal{Z}}_j)$ vanish identically; for example,
\begin{equation}
    \begin{split}
    p_jh_{j+2}p_j
    &=h_{j+2}\frac{1}{2}\left(\bar{\mathcal{I}}_j-x_j\bar{\mathcal{X}}_j-y_j\bar{\mathcal{Y}}_j+z_j\bar{\mathcal{Z}}_j\right)p_j=0.
    \end{split}
\end{equation}
Similarly we find
\begin{equation}
    p_jh_{j+1}h_{j+2}h_{j+3}p_j=p_j\mathrm{i}h_{j+1}h_{j+2}p_j=p_j\mathrm{i}h_{j+2}h_{j+3}p_j=0.
\end{equation}
Thus $e_j e_{j+1} e_j$ is calculated as
\begin{equation}
    \begin{split}
    e_j e_{j+1} e_j&=\Delta^3 p_{2j}p_{2j+2}p_{2j}\\
    &=\frac{\Delta^3}{4}\left(1-z_{2j}h_{2j+3}-z_{2j}z_{2j+2}+z_{2j+2}h_{2j+3}\right)p_{2j}\\
    &=\frac{\Delta^2}{4}\left((1-z_{2j}z_{2j+2})-(z_{2j}-z_{2j+2})h_{2j+3}\right)e_j 
    \end{split}
\end{equation}
For the right-hand side to be equal to $e_j$, the parameters $(\Delta,z_{2j},z_{2j+2})$ need to satisfy the following relations:
\begin{equation}
    \frac{{\Delta}^2}{4}(1-z_{2j}z_{2j+2})=1,~~z_{2j}-z_{2j+2}=0,
\end{equation}
or
\begin{equation}
\label{eq:Delta-z-q}
    \Delta=q+q^{-1},~~z_{2j}=z_{2j+2}=\frac{q-q^{-1}}{q+q^{-1}},
\end{equation}
where $q$ is a real parameter.
The other gluing condition $e_je_{j-1}e_j$ leads to the same relations (\ref{eq:Delta-z-q}).
The two parameters $x_{2j}$ and $y_{2j}$ are freely chosen on the circle of radius $2/(q+q^{-1})$ on each $j$.
Therefore, we conclude that the TL generators $e_j$ consisting of the three consecutive vertices are obtained as
\begin{equation}
\label{eq:TL-generator-on-3}
\begin{split}
    &e_j=\frac{q+q^{-1}}{2}\left( \mathcal{I}_{2j}+x_{2j}\mathcal{X}_{2j} +y_{2j}\mathcal{Y}_{2j}+\frac{q-q^{-1}}{q+q^{-1}}\mathcal{Z}_{2j}\right),\\
    &\hspace{20mm} x_{2j}^2 + y_{2j}^2 = \left(\frac{2}{q+q^{-1}}\right)^2,
\end{split}
\end{equation}
which is our main result.

Finally, we refer to the case of $\bar{p_j}$ ($\ref{eq:projector-on-3-conj}$).
The same steps of the case of $p_j$ (\ref{eq:projector-on-3}), leads to the following conditions instead of  (\ref{eq:Delta-z-q}),
\begin{equation}
    \Delta=q+q^{-1},~~z_{2j}=-z_{2j+2}=\frac{q-q^{-1}}{q+q^{-1}}.
\end{equation}
Thus the TL generators $\{e_{j}\}_{j=\mathrm{odd}}$ have the similar form of (\ref{eq:TL-generator-on-3}):
\begin{equation}
\begin{split}
    &\bar{e}_{j=\mathrm{odd}}=\frac{q+q^{-1}}{2}\left( \bar{\mathcal{I}}_{2j}+x_{2j}\bar{\mathcal{X}}_{2j} +y_{2j}\bar{\mathcal{Y}}_{2j}+\frac{q-q^{-1}}{q+q^{-1}}\bar{\mathcal{Z}}_{2j}\right),\\
    &\hspace{20mm} x_{2j}^2 + y_{2j}^2 = \left(\frac{2}{q+q^{-1}}\right)^2,
\end{split}
\end{equation}
but $\{e_{j}\}_{j=\mathrm{even}}$ have the form of the reverse of $z_{2j}$:
\begin{equation}
\begin{split}
    &\bar{e}_{j=\mathrm{even}}=\frac{q+q^{-1}}{2}\left( \bar{\mathcal{I}}_{2j}+x_{2j}\bar{\mathcal{X}}_{2j} +y_{2j}\bar{\mathcal{Y}}_{2j}-\frac{q-q^{-1}}{q+q^{-1}}\bar{\mathcal{Z}}_{2j}\right),\\
    &\hspace{20mm} x_{2j}^2 + y_{2j}^2 = \left(\frac{2}{q+q^{-1}}\right)^2.
\end{split}
\end{equation}

\subsection{Representations of the constructed TL generators}

In this subsection, we discuss representations of the constructed TL generators.
Using the TFI representation (\ref{eq:1dTFI-Rep1}) of $\{h_j\}$, we find the pseudo Pauli operators in this representation are given as
\begin{equation}
    \begin{split}
    &\rho_{\mathrm{TFI}}\left(\mathcal{I}_{2j}\right)=\frac{1}{2}\left(1-Z_jZ_{j+1}\right), ~~
    \rho_{\mathrm{TFI}}\left(\mathcal{X}_{2j}\right) =\frac{1}{2}\left(X_jX_{j+1}+Y_jY_{j+1}\right),\\
    &\rho_{\mathrm{TFI}}\left(\mathcal{Y}_{2j}\right)=\frac{1}{2}\left(-Y_jX_{j+1}+X_jY_{j+1}\right), ~~
    \rho_{\mathrm{TFI}}\left(\mathcal{Z}_{2j}\right) =\frac{1}{2}\left(Z_j-Z_{j+1}\right).\\
    \end{split}
\end{equation}
Thus in the case of $(x_{2j}, y_{2j})=(-2/(q+q^{-1}), 0)$, the above representation of the TL generators coincide with the local Hamiltonian of the spin-1/2 XXZ model;
\begin{equation}
    \begin{split}
        \rho_{\mathrm{TFI}}\left(e_j\right)
        &=-\mathcal{X}_{2j}+\frac{q+q^{-1}}{2}\mathcal{I}_{2j}+\frac{q+q^{-1}}{2}\mathcal{Z}_{2j}\\
        &=-\frac{1}{2}\left(X_jX_{j+1}+Y_jY_{j+1}\right)
        -\frac{q+q^{-1}}{4}\left(Z_jZ_{j+1}-1\right)
        +\frac{q-q^{-1}}{4}\left(Z_j-Z_{j+1}\right)
    \end{split}
\end{equation}

There are other representations for the bond generators ${h_j}$.
The first example is the alternating-XY representation:
\begin{equation}
    \rho_{\mathrm{AXY}}\left(h_{2j-1}\right)=X_{2j-1}X_{2j},~~\rho_{\mathrm{AXY}}\left(h_{2j}\right)=Y_{2j}Y_{2j+1}.
\end{equation}
In this case, the corresponding TL spin system is given as
\begin{equation}
    \begin{split}
    &\rho_{\mathrm{AXY}}\left(\mathcal{I}_{2j}\right)=\frac{1}{2}\left(1-X_{2j-1}X_{2j}X_{2j+1}X_{2j+2}\right), \\
    &\rho_{\mathrm{AXY}}\left(\mathcal{X}_{2j}\right) =\frac{1}{2}\left(Y_{2j}Y_{2j+1}+X_{2j-1}Z_{2j}Z_{2j+1}X_{2j+2}\right),\\
    &\rho_{\mathrm{AXY}}\left(\mathcal{Y}_{2j}\right)=\frac{1}{2}\left(-X_{2j-1}Z_{2j}Y_{2j+1}+Y_{2j}Z_{2j+1}X_{2j+2}\right), \\
    &\rho_{\mathrm{AXY}}\left(\mathcal{Z}_{2j}\right) =\frac{1}{2}\left(X_{2j-1}X_{2j}-X_{2j+1}X_{2j+2}\right).\\
    \end{split}
\end{equation}
The second term of $\rho_{\mathrm{AXY}}\left(\mathcal{I}_{2j}\right)$ is the four-body spin interaction.
The cluster interactions are emerged in the second term of $\rho_{\mathrm{AXY}}\left(\mathcal{X}_{2j}\right)$ and the first and second terms of $\rho_{\mathrm{AXY}}\left(\mathcal{Y}_{2j}\right)$.

The second example is the cluster representation:
\begin{equation}
    \rho_{\mathrm{XZX}}\left(h_{2j-1}\right)=X_{2j-1}Z_{2j}X_{2j+1},~~\rho_{\mathrm{XZX}}\left(h_{2j}\right)=X_{2j}X_{2j+2}.
\end{equation}
In this case, the corresponding TL spin system is given as
\begin{equation}
    \begin{split}
    &\rho_{\mathrm{XZX}}\left(\mathcal{I}_{2j}\right)=\frac{1}{2}\left(1-X_{2j-1}Z_{2j}Z_{2j+2}X_{2j+3}\right), \\
    &\rho_{\mathrm{XZX}}\left(\mathcal{X}_{2j}\right) =\frac{1}{2}\left(X_{2j}X_{2j+2}+X_{2j-1}Y_{2j}Y_{2j+2}X_{2j+3}\right),\\
    &\rho_{\mathrm{XZX}}\left(\mathcal{Y}_{2j}\right)=\frac{1}{2}\left(-X_{2j-1}Y_{2j}X_{2j+1}X_{2j+2}+X_{2j}X_{2j+1}Y_{2j+2}X_{2j+3}\right), \\
    &\rho_{\mathrm{XZX}}\left(\mathcal{Z}_{2j}\right) =\frac{1}{2}\left(X_{2j-1}Z_{2j}X_{2j+1}-X_{2j+1}Z_{2j+2}X_{2j+3}\right).
    \end{split}
\end{equation}

Finally, we consider Majorana representations of these models and the degeneracies of the spectrums.
From the knowledge of the paper \cite{Ogura-Imamura-Kameyama-Minami-Sato-2020}, each $h_j$ is represented by a Majorana bilinear: $h_j=\mathrm{i}\epsilon_j\varphi_{2j-1}\varphi_{2j}$.
The factor $\epsilon_j$s are different depending on the spin representations $\rho$.
We use this transformation for the bond generators, and then the Hamiltonians with the TL algebra has the same forms as Majorana fermions.
Thus the Hamiltonians have the same spectrums but the degeneracies are different depending on the spin representations.
In the case of $\rho_{\mathrm{TFI}}$ and $\rho_{\mathrm{AXY}}$, the dimension of Hilbert space of spins is the same as that of the Majorana fermions, and therefore we can choose the factor $\epsilon_j$ freely as $1$:
\begin{equation}
    \rho_{\mathrm{TFI}}(h_j)=\rho_{\mathrm{AXY}}(h_j) = \mathrm{i} \varphi_{2j-1}\varphi_{2j}.
\end{equation}
In the case of $\rho_{\mathrm{XZX}}$, the factor $\epsilon_j$ becomes a local conserved charge operator which takes a value of $\pm1$, because the dimension of Hilbert space of spins is twice larger than that of the Majorana fermions; consequently
\begin{equation}
    \rho_{\mathrm{XZX}}(h_j)= \mathrm{i} \epsilon_j\varphi_{2j-1}\varphi_{2j}.
\end{equation}
We also see this fact from the usual Jordan-Wigner transformation:
\begin{equation}
X_j=\Tilde{\varphi}_{2j-1} \prod_{k<j} \mathrm{i} \Tilde{\varphi}_{2k-1}\Tilde{\varphi}_{2k},~~
Z_j=\mathrm{i} \Tilde{\varphi}_{2j-1}\Tilde{\varphi}_{2j}.
\end{equation}
Here $\Tilde{\varphi}$ is Majorana fermion.
Using this transformations, we find that
\begin{equation}
\label{eq:4-Majorana}
\begin{split}
    &\rho_{\mathrm{XZX}}\left(h_{2j-1}\right)=X_{2j-1}Z_{2j}X_{2j+1}
    =\mathrm{i}\Tilde{\varphi}_{4j-2}\Tilde{\varphi}_{4j+1},\\
    &\rho_{\mathrm{XZX}}\left(h_{2j}\right)=X_{2j}X_{2j+2}
    =\Tilde{\varphi}_{4j}\Tilde{\varphi}_{4j+1}\Tilde{\varphi}_{4j+2}\Tilde{\varphi}_{4j+3}.
\end{split}
\end{equation}
The combination of the two Majorana fermions, $\epsilon_j=\mathrm{i}\Tilde{\varphi}_{4j}\Tilde{\varphi}_{4j+3}$, commute with every term of $\rho_{\mathrm{XZX}}\left(h_{2j-1}\right)$ and $\rho_{\mathrm{XZX}}\left(h_{2j}\right)$, and therefore the four products of Majorana fermions, which is the second line of (\ref{eq:4-Majorana}), is deformed into the product of a conserved charge and a Majorana bilinear:
\begin{equation}
    \Tilde{\varphi}_{4j}\Tilde{\varphi}_{4j+1}\Tilde{\varphi}_{4j+2}\Tilde{\varphi}_{4j+3}=-\mathrm{i}\epsilon_j\Tilde{\varphi}_{4j+1}\Tilde{\varphi}_{4j+2}.
\end{equation}
This is equivalent to the aforementioned result.
We can conclude that the constructed spin systems have the same spectrum with the $S=1/2$ XXZ spin system, but the degeneracy is different depending on the representations.


\section{Conclusion}
\label{sec:conclusion}

In this paper, we have explicitly constructed the TL algebra by the generators of the 1dTFI bond algebra on one, two, and three consecutive vertices.
When we take the Pauli-matrix representation (i.e. the 1dTFI model itself) of the generator of the 1dTFI bond algebra, then the TL generators constructed here coincide with the local Hamiltonian of the $S=1/2$ XXZ model.
Furthermore, we derive other integrable spin systems from other spin representations and discuss the spectrums and degeneracy.
This illustrates the connection between the class whose Hamiltonian can be written as Majorana bilinears and the class associated with the quantum integrability.

Our method may reveal interesting aspects of bond algebras and lattice spin systems.
It is known that The $S=1/2$ XXZ model has the quantum group symmetry $U_q(sl_2)$, where $q$ is the deformation parameter \cite{Pasquier-Saleur-1990}.
Each TL generator that we constructed has the parameter $q$, which we have introduced from the TL gluing condition.
Investing in how to express and understand the quantum group symmetry from the perspective of the bond algebra is left for future work.

We have only considered the 1dTFI bond algebra.
We can think of other cases such as the Kitaev honeycomb model \cite{Minami-2019,Ogura-Imamura-Kameyama-Minami-Sato-2020}.
Generalization of the TL algebra to higher dimensions has not yet been studied well.
Extension of $S=1/2$ to general $S$ is also an important work.
The TL algebra of $S=1$ was discussed in \cite{Barber-Batchelor-1989,Batchelor-Mezincescu-Nepomechie-Rittenberg-1990}.
Our algebraic approach provides a systematic way to explore other interesting lattice systems.

\newpage
\section*{Acknowledgment}

I thank Keisuke Totsuka and Masatoshi Sato for several helpful comments about this work.


%

\vspace{0.2cm}


\let\doi\relax


\bibliographystyle{ptephy.bst}
\bibliography{TFI-TL.bib}

\appendix

\section{Renormalization group of $\mathcal{BA}_\mathrm{1dTFI}$}
\label{appendix:RG}

In this appendix, we consider a renormalization group of the 1dTFI model.
Following the previous study \cite{Monthus-2015}, we use the projection operators constructed in Sec. \ref{subsec:TL-on-2}.

\begin{figure}
    \centering
    \includegraphics[width=12cm]{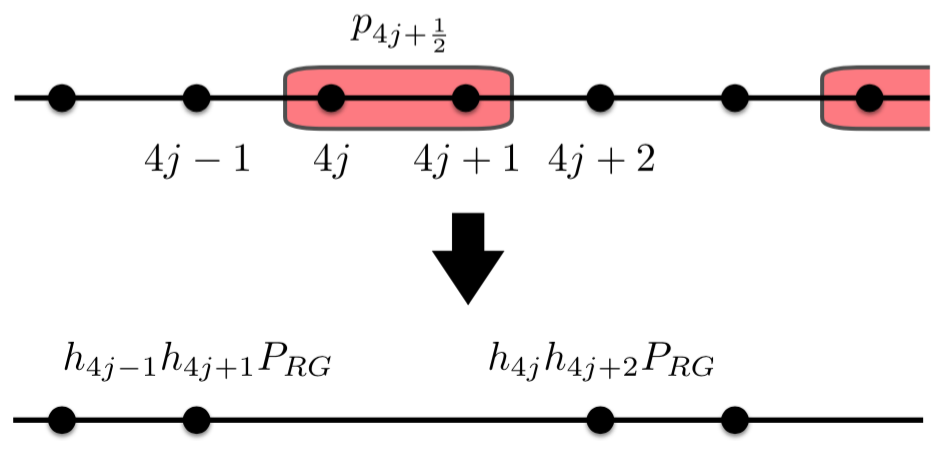}
    \caption{Renormalization of the 1dTFIM bond algebra. Projection operators are defined on two consecutive vertices $4j$ and $4j+1$. The renormalized graph has only the half number of vertices as the original one.}
    \label{fig:RG}
\end{figure}

The 1dTFI Hamiltonian (\ref{eq:1dTHI-Hamiltonian}) is written by using $\{h_j\}$ as follows:
\begin{equation}
\begin{split}
    H
    &=\sum_j (\lambda_{\mathrm{odd}} h_{2j-1} + \lambda_{\mathrm{even}} h_{2j})\\
    &=\sum_j H_{4j+\frac{1}{2}},
\end{split}
\end{equation}
\begin{equation}
\label{eq:local-Hamiltonian}
    H_{4j+\frac{1}{2}}=
    \lambda_{\mathrm{odd}} h_{4j-1} + 
    \lambda_{\mathrm{even}} h_{4j}+
    \lambda_{\mathrm{odd}} h_{4j+1} + 
    \lambda_{\mathrm{even}} h_{4j+2}.
\end{equation}
We pick up the second and third term in (\ref{eq:local-Hamiltonian}), and consider the projection of its ground state.
Since
\begin{equation}
    (\lambda_{\mathrm{even}} h_{4j}+
    \lambda_{\mathrm{odd}} h_{4j+1})^2=\lambda_{\mathrm{even}}^2+
    \lambda_{\mathrm{odd}}^2,
\end{equation}
then the operator which project the Hilbert space into the ground state of the local operators $\lambda_{\mathrm{even}} h_{4j}+\lambda_{\mathrm{odd}} h_{4j+1}$ is
\begin{equation}
    p_{4j+\frac{1}{2}}=
    \frac{1}{2}\left(
    1+
    \frac{\lambda_{\mathrm{even}}}
    {\sqrt{\lambda_{\mathrm{even}}^2+\lambda_{\mathrm{odd}}^2}} h_{4j}+
    \frac{\lambda_{\mathrm{odd}}}
    {\sqrt{{\lambda_{\mathrm{even}}^2+\lambda_{\mathrm{odd}}^2}}} h_{4j+1}
    \right).
\end{equation}
We find that this is equivalent to the projection operator whose parameters are $(x,y,z)=(\lambda_{\mathrm{even}}/\sqrt{\lambda_{\mathrm{even}}^2+
    \lambda_{\mathrm{odd}}^2},\lambda_{\mathrm{odd}}/\sqrt{{\lambda_{\mathrm{even}}^2+
    \lambda_{\mathrm{odd}}^2}},0)$.
Let us define a renormalization projector as
\begin{equation}
    P_{\mathrm{RG}}=\prod_j p_{4j+\frac{1}{2}}.
\end{equation}
The application of this to the local Hamiltonian (\ref{eq:local-Hamiltonian}) are computed as
\begin{equation}
    \begin{split}
    &P_{\mathrm{RG}}H_{4j+\frac{1}{2}}P_{\mathrm{RG}}\\
    &=
    \left[p_{4j+\frac{1}{2}}
    (\lambda_{\mathrm{odd}} h_{4j-1} + 
    \lambda_{\mathrm{even}} h_{4j+2})
    p_{4j+\frac{1}{2}}
    + p_{4j+\frac{1}{2}}
    (\lambda_{\mathrm{even}} h_{4j}+
    \lambda_{\mathrm{odd}} h_{4j+1})
    p_{4j+\frac{1}{2}}\right]\prod_{k\neq j}p_{4k+\frac{1}{2}}\\
    &=
    \left[
    (\lambda_{\mathrm{odd}}y h_{4j-1}h_{4j+1} + 
    \lambda_{\mathrm{even}}x h_{4j}h_{4j+2})
    +
    (\lambda_{\mathrm{even}}x+
    \lambda_{\mathrm{odd}} y)
    \right]\prod_{j}p_{4j+\frac{1}{2}}\\
    &=
    \left(H^{\mathrm{R}}_{4j+\frac{1}{2}} + \sqrt{{\lambda_{\mathrm{even}}^2+\lambda_{\mathrm{odd}}^2}}
    \right)P_{\mathrm{RG}}.
    \end{split}
\end{equation}
Here the renormalized local Hamiltonian $H^{\mathrm{R}}_{4j+\frac{1}{2}}$ has the form
\begin{equation}
    H^{\mathrm{R}}_{4j+\frac{1}{2}} = \frac{\lambda_{\mathrm{odd}}^2}
    {\sqrt{{\lambda_{\mathrm{even}}^2+\lambda_{\mathrm{odd}}^2}}} h_{4j-1}h_{4j+1} + 
    \frac{\lambda_{\mathrm{even}}^2}
    {\sqrt{{\lambda_{\mathrm{even}}^2+\lambda_{\mathrm{odd}}^2}}} h_{4j}h_{4j+2},
\end{equation}
and commute with the renormalization projector $P_{\mathrm{RG}}$.
The two terms in $H^{\mathrm{R}}_{4j+\frac{1}{2}}$ anticommute with each other.
Thus, the renormalized graph of $H^R=\sum_j H^{\mathrm{R}}_{4j+\frac{1}{2}}$ has the same form of the original $H$, but the number of vertices is half.
The renormalized coupling constants are

\begin{equation}
\label{eq:lambdaRG}
    \lambda^R_{\mathrm{odd}}=
    \frac{\lambda_{\mathrm{odd}}^2}
    {\sqrt{{\lambda_{\mathrm{even}}^2+\lambda_{\mathrm{odd}}^2}}},
    ~~
    \lambda^R_{\mathrm{even}}=
    \frac{\lambda_{\mathrm{even}}^2}
    {\sqrt{{\lambda_{\mathrm{even}}^2+\lambda_{\mathrm{odd}}^2}}}.
\end{equation}
Using $K=\lambda_{\mathrm{odd}}/\lambda_{\mathrm{even}}$, (\ref{eq:lambdaRG}) is summarized as $K^R=K^2$.
Thus, the disordered attractive fixed point $K = 0$ and the ferromagnetic attractive fixed point $K \to +\infty$ are separated by the unstable fixed point $K_C = 1$.
This result is equivalent to Monthus's work \cite{Monthus-2015}.

\end{document}